\documentclass[superscriptaddress,aps,prl,amsmath,amssymb,bibnotes,altaffilletter,twocolumn]{revtex4-2}

\usepackage{hyperref}
\usepackage[T1]{fontenc}
\usepackage{xcolor}
\usepackage{graphicx}
\usepackage{dcolumn}
\usepackage{bm}

\hypersetup{colorlinks=true, citecolor=blue, urlcolor=blue, linkcolor=blue}

\begin{document}
\def\nuc#1#2{${}^{#1}$#2}
\def\mee{$\langle m_{\beta\beta} \rangle$}
\def\mnu{$m_{\nu}$}
\def\ml{$m_{lightest}$}
\def\gnu{$\langle g_{\nu,\chi}\rangle$}
\def\mmod{$\| \langle m_{\beta\beta} \rangle \|$}
\def\mb{$\langle m_{\beta} \rangle$}
\def\BBz{$\beta\beta(0\nu)$}
\def\BBm{$\beta\beta(0\nu,\chi)$}
\def\BBt{$\beta\beta(2\nu)$}
\def\nonubb{$\beta\beta(0\nu)$}
\def\twonubb{$\beta\beta(2\nu)$}
\def\BB{$\beta\beta$}
\def\Mz{$M_{0\nu}$}
\def\Mt{$M_{2\nu}$}
\def\MzG{$M^{GT}_{0\nu}$}           
\def\MzF{$M^{F}_{0\nu}$}                
\def\MtG{$M^{GT}_{2\nu}$}           
\def\MtF{$M^{F}_{2\nu}$}                
\def\Gz{$G_{0\nu}$}					
\def\Tz{$T^{0\nu}_{1/2}$}
\def\Tt{$T^{2\nu}_{1/2}$}
\def\Tc{$T^{0\nu\,\chi}_{1/2}$}
\def\Rz{$\Gamma_{0\nu}$}            
\def\Rt{$\Gamma_{2\nu}$}            
\def\ms{$\delta m_{\rm sol}^{2}$}
\def\ma{$\delta m_{\rm atm}^{2}$}
\def\mot{$\delta m_{12}^{2}$}
\def\mtt{$\delta m_{23}^{2}$}
\def\ts{$\theta_{\rm sol}$}
\def\ta{$\theta_{\rm atm}$}
\def\ttwo{$\theta_{12}$}
\def\tot{$\theta_{13}$}
\def\gpp{$g_{pp}$}                  
\def\gA{$g_{A}$}                  
\def\qval{$Q_{\beta\beta}$}                 
\def\be{\begin{equation}}
\def\ee{\end{equation}}
\def\cpKkgy{cnts/(keV kg y)}
\def\cpKkgd{cnts/(keV kg d)}
\def\cpRty{cnts/(ROI t y)}
\def\onecpRty{1~cnt/(ROI t y)}
\def\threecpRty{3~cnts/(ROI t y)}
\def\ppc{P-PC}                          
\def\nsc{N-SC}                          
\def\cosixty{$^{60}Co$}
\def\thttt{$^{232}\mathrm{Th}$}
\def\utte{$^{238}\mathrm{U}$}
\def\mubqkg{$\mu\mathrm{Bq/kg}$}
\def\cusulfate{$\mathrm{CuSO}_4$}
\def\MJ{{\sc Majorana}}             
\def\DEM{{\sc Demonstrator}}             
\def\MJDEMbf{\bfseries{\scshape{Majorana Demonstrator}}}
\def\MJbf{\bfseries{\scshape{Majorana}}}
\def\MJDEMit{\itshape{\scshape{Majorana Demonstrator}}}
\newcommand{\Gerda}{GERDA}
\newcommand{\GF}{\textsc{Geant4}}
\newcommand{\MaGe}{\textsc{MaGe}}
\def\znbb{$0\nu\beta\beta$}

\newcommand{\bl}[1]{\textcolor{blue}{#1}} 

\preprint{APS/123-QED}

\title{A Search for Spontaneous Radiation from Wavefunction Collapse \\ in the \textsc{Majorana Demonstrator}}

\newcommand{\ITEP}{National Research Center ``Kurchatov Institute'' Institute for Theoretical and Experimental Physics, Moscow, 117218 Russia}
\newcommand{\JINR}{Joint Institute for Nuclear Research, Dubna, 141980 Russia} 
\newcommand{\lbnl}{Nuclear Science Division, Lawrence Berkeley National Laboratory, Berkeley, CA 94720, USA}
\newcommand{\lbnle}{Engineering Division, Lawrence Berkeley National Laboratory, Berkeley, CA 94720, USA}
\newcommand{\lanl}{Los Alamos National Laboratory, Los Alamos, NM 87545, USA}
\newcommand{\queens}{Department of Physics, Engineering Physics and Astronomy, Queen's University, Kingston, ON K7L 3N6, Canada}
\newcommand{\uw}{Center for Experimental Nuclear Physics and Astrophysics, and Department of Physics, University of Washington, Seattle, WA 98195, USA}
\newcommand{\unc}{Department of Physics and Astronomy, University of North Carolina, Chapel Hill, NC 27514, USA}
\newcommand{\duke}{Department of Physics, Duke University, Durham, NC 27708, USA}
\newcommand{\ncsu}{Department of Physics, North Carolina State University, Raleigh, NC 27695, USA}	
\newcommand{\ornl}{Oak Ridge National Laboratory, Oak Ridge, TN 37830, USA}
\newcommand{\ou}{Research Center for Nuclear Physics, Osaka University, Ibaraki, Osaka 567-0047, Japan}
\newcommand{\pnnl}{Pacific Northwest National Laboratory, Richland, WA 99354, USA}
\newcommand{\ttu}{Tennessee Tech University, Cookeville, TN 38505, USA}
\newcommand{\sdsmt}{South Dakota Mines, Rapid City, SD 57701, USA}
\newcommand{\usc}{Department of Physics and Astronomy, University of South Carolina, Columbia, SC 29208, USA}
\newcommand{\usd}{Department of Physics, University of South Dakota, Vermillion, SD 57069, USA}  
\newcommand{\ut}{Department of Physics and Astronomy, University of Tennessee, Knoxville, TN 37916, USA}
\newcommand{\tunl}{Triangle Universities Nuclear Laboratory, Durham, NC 27708, USA}
\newcommand{\mpi}{Max-Planck-Institut f\"{u}r Physik, M\"{u}nchen, 80805, Germany}
\newcommand{\tum}{Physik Department and Excellence Cluster Universe, Technische Universit\"{a}t, M\"{u}nchen, 85748 Germany}
\newcommand{\williams}{Physics Department, Williams College, Williamstown, MA 01267, USA}
\newcommand{\ciemat}{Centro de Investigaciones Energ\'{e}ticas, Medioambientales y Tecnol\'{o}gicas, CIEMAT 28040, Madrid, Spain}
\newcommand{\iu}{Department of Physics, Indiana University, Bloomington, IN 47405, USA}
\newcommand{\iuceem}{IU Center for Exploration of Energy and Matter, Bloomington, IN 47408, USA}

\author{I.J.~Arnquist}\affiliation{\pnnl} 
\author{F.T.~Avignone~III}\affiliation{\usc}\affiliation{\ornl}
\author{A.S.~Barabash}\affiliation{\ITEP}
\author{C.J.~Barton}\affiliation{\usd}	
\author{K.H.~Bhimani}\affiliation{\unc}\affiliation{\tunl}
\author{E.~Blalock}\affiliation{\ncsu}\affiliation{\tunl} 
\author{B.~Bos}\affiliation{\unc}\affiliation{\tunl} 
\author{M.~Busch}\affiliation{\duke}\affiliation{\tunl}	
\author{M.~Buuck}\altaffiliation{SLAC National Accelerator Laboratory, Menlo Park, CA 94025, USA}\affiliation{\uw}
\author{T.S.~Caldwell}\affiliation{\unc}\affiliation{\tunl}	
\author{Y-D.~Chan}\affiliation{\lbnl}
\author{C.D.~Christofferson}\affiliation{\sdsmt} 
\author{P.-H.~Chu}\affiliation{\lanl} 
\author{M.L.~Clark}\affiliation{\unc}\affiliation{\tunl} 
\author{C.~Cuesta}\affiliation{\ciemat}	
\author{J.A.~Detwiler}\affiliation{\uw}	
\author{Yu.~Efremenko}\affiliation{\ut}\affiliation{\ornl}
\author{H.~Ejiri}\affiliation{\ou}
\author{S.R.~Elliott}\affiliation{\lanl}
\author{G.K.~Giovanetti}\affiliation{\williams}  
\author{M.P.~Green}\affiliation{\ncsu}\affiliation{\tunl}\affiliation{\ornl}   
\author{J.~Gruszko}\affiliation{\unc}\affiliation{\tunl} 
\author{I.S.~Guinn}\affiliation{\unc}\affiliation{\tunl} 
\author{V.E.~Guiseppe}\affiliation{\ornl}	
\author{C.R.~Haufe}\affiliation{\unc}\affiliation{\tunl}	
\author{R.~Henning}\affiliation{\unc}\affiliation{\tunl}
\author{D.~Hervas~Aguilar}\affiliation{\unc}\affiliation{\tunl} 
\author{E.W.~Hoppe}\affiliation{\pnnl}
\author{A.~Hostiuc}\affiliation{\uw} 
\author{I.~Kim}~\email{inwookkim@lanl.gov}\affiliation{\lanl} 
\author{R.T.~Kouzes}\affiliation{\pnnl}
\author{T.E.~Lannen~V}\affiliation{\usc} 
\author{A.~Li}\affiliation{\unc}\affiliation{\tunl}
\author{A.M.~Lopez}\affiliation{\ut}	
\author{J.M. L\'opez-Casta\~no}\affiliation{\ornl} 
\author{E.L.~Martin}\affiliation{\unc}\affiliation{\tunl}	
\author{R.D.~Martin}\affiliation{\queens}	
\author{R.~Massarczyk}\affiliation{\lanl}		
\author{S.J.~Meijer}\affiliation{\lanl}	
\author{T.K.~Oli}\affiliation{\usd}  
\author{G.~Othman}\altaffiliation{Institute for Experimental Physics; Universit{\"a}t Hamburg; Luruper Chaussee 149; 22761 Hamburg, Germany}\affiliation{\unc}\affiliation{\tunl} 
\author{L.S.~Paudel}\affiliation{\usd} 
\author{W.~Pettus}\affiliation{\iu}\affiliation{\iuceem}	
\author{A.W.P.~Poon}\affiliation{\lbnl}
\author{D.C.~Radford}\affiliation{\ornl}
\author{A.L.~Reine}\affiliation{\unc}\affiliation{\tunl}	
\author{K.~Rielage}\affiliation{\lanl}
\author{N.W.~Ruof}\affiliation{\uw}	
\author{D.~Tedeschi}\affiliation{\usc}		
\author{R.L.~Varner}\affiliation{\ornl}  
\author{S.~Vasilyev}\affiliation{\JINR}	
\author{J.F.~Wilkerson}\affiliation{\unc}\affiliation{\tunl}\affiliation{\ornl}    
\author{C.~Wiseman}\affiliation{\uw}		
\author{W.~Xu}\affiliation{\usd} 
\author{C.-H.~Yu}\affiliation{\ornl}
\author{B.X.~Zhu}\altaffiliation{Present address: Jet Propulsion Laboratory, California Institute of Technology, Pasadena, CA 91109, USA}\affiliation{\lanl} 

\collaboration{{\sc{Majorana}} Collaboration}
\noaffiliation

\date{\today}

\begin{abstract}
    The \MJ{} \DEM{} neutrinoless double-beta decay experiment comprises a 44~kg~(30~kg enriched in $^{76}$Ge) array of p-type, point-contact germanium detectors. 
    With its unprecedented energy resolution and ultra-low backgrounds, \MJ{} also searches for rare event signatures from beyond Standard Model physics in the low energy region below 100~keV.
    In this letter, we test the continuous spontaneous localization~(CSL) model, one of the mathematically well-motivated wavefunction collapse models aimed at solving the long-standing unresolved quantum mechanical measurement problem. 
    While the CSL predicts the existence of a detectable radiation signature in the X-ray domain, we find no evidence of such radiation in the 19-100~keV range in a 37.5~kg-y enriched germanium exposure collected between Dec.~31, 2015 and Nov.~27, 2019 with the \DEM{}.
    We explored both the non-mass-proportional~(n-m-p) and the mass-proportional~(m-p) versions of the CSL with two different assumptions: that only the quasi-free electrons can emit the X-ray radiation and that the nucleus can coherently emit an amplified radiation.
    In all cases, we set the most stringent upper limit to date for the white CSL model on the collapse rate, $\lambda$, providing a factor of 40-100 improvement in sensitivity over comparable searches.  
    Our limit is the most stringent for large parts of the allowed parameter space.
    If the result is interpreted in terms of the Di\`{o}si-Penrose~(DP) gravitational wavefunction collapse model, the lower bound with a 95\% CL confidence level is almost an order of magnitude improvement over the previous best limit.
\end{abstract}

\maketitle

Spontaneous wavefunction collapse models~\cite{ghirardi1986,ghirardi1987,ghirardi1988puzzling,ghirardi1990old,bassi2003models,Bassi2013models,Ferialdi2012,diosi1989models,penrose1996gravity} aim at solving the long-standing unresolved measurement problem of quantum mechanics through a stochastic nonlinear modification of the Schr\"{o}dinger equation. 
The additional phenomenological term in the Schr\"{o}dinger equation, which fundamentally breaks the quantum superposition principle in large scale systems, is interpreted as a universal noise field defined at each point in spacetime~\cite{Bassi_2016}.
In these models, the interaction with the noise field localizes a system even in the absence of a measurement process, always resulting in definite outcomes after a quantum mechanical measurement.
The rate of the wavefunction collapse scales with the size of the system, resulting in a fast localization of macroscopic systems while not significantly affecting the standard quantum mechanical motion of the microscopic world.
These collapse models are considered to be one of the few mathematically consistent, testable quantum theories of wavefunction collapse~\cite{Bassi2013models,Curceanu_2015,vinante2020layered}. 

The continuous spontaneous radiation~(CSL) model is one of the most well-studied among collapse models. 
A series of theoretical papers~\cite{ghirardi1986,ghirardi1987,ghirardi1988puzzling,ghirardi1990old,pearle1989combining,Adler_2007_bound,Adler_2013_spontaneous} have developed a consistent theory, in which particles undergo spontaneous localization around definite positions following a Poisson distribution characterized by a mean frequency $\lambda$, which is the collapse rate, and a correlation length $r_\mathrm{C}$, which is the spatial resolution of the collapse. 
Larger values of $\lambda$~(i.e., faster localization) imply the quantum-to-classical transition would occur at smaller mesoscopic scales, while smaller $\lambda$ values~(i.e., slower localization) indicate the transitions occur at larger, macroscopic scales. 
The paper by Adler~\cite{Adler_2007_bound} provides a summary of the various experimental limits on the theory. 

One consequence of the CSL is that a free charged particle will be accelerated during the collapse and will emit electromagnetic radiation, which is not predicted by standard quantum mechanics~\cite{fu1997quantum,Adler_2007_bound,Adler_2013_spontaneous,donadi2014emission,donadi2014spntaneous,piscicchia2017igex,donadi2021novel}. 
Detection of this radiation in the X-ray regime would be a direct test of the CSL theory. 
In the na\"{i}ve version of the CSL, where the collapse noise is modelled as white noise, the measurement of X-ray radiation between $10^{16}-10^{20}$~Hz~($0.1-100$~keV) can place a stringent limit on the model. 
A colored noise CSL model was later introduced, motivated by the claim that any field with a physical origin should always have a non-flat spectrum with a cut-off frequency~($\Omega$)~\cite{Pearle1993,Adler_2007collapse,Adler_2008collapse,Adler_2013_spontaneous,toros2017colored}. 
In these colored extensions, the value of $\Omega$ is commonly chosen to be $10^{11}-10^{12}$~Hz, similar to some of the most common backgrounds with cosmological origins~\cite{toros2017colored}. 
The radiation in the X-ray regime exceeds this cut-off, and therefore the constraints from X-ray measurements become weaker in the colored extensions~\cite{carlesso2018colored}. 
A CSL signature search using X-ray measurements does not only set a stringent limit on previously unexplored parameter space, but also serves as a direct test of the white CSL model. 
In this letter, we only consider the case of the white CSL.

The rate of X-rays emitted by a free particle in the CSL model was first calculated by Fu~\cite{fu1997quantum} under two different assumptions on the coupling $\alpha_\mathrm{c}$ with the noise field: the non-mass-proportional~(n-m-p) version where $\alpha_\mathrm{c}$ is thought to be independent of the particle mass, and the mass-proportional~(m-p) version where  $\alpha_\mathrm{c}$ is considered to be proportional to the particle mass.
We examine both cases to set the most stringent limits.

We first consider the n-m-p version of the CSL. 
Since the Bremsstrahlung radiation from an accelerated charged particle of mass $m_\mathrm{X}$ is proportional to  $\alpha_\mathrm{c}^2/m_\mathrm{X}^2$, the X-ray emission rate in this case is suppressed by $m_\mathrm{X}^2$. 
A spontaneous radiation emitted from acceleration of a free particle would have an energy~$(E)$ dependence of $1/E$, with the emission rate $d\Gamma(E)$ given by

\begin{equation}\label{eq:radiation_rate_nmp}
    \frac{d\Gamma(E)}{dE} = A_f \times \frac{\hbar \lambda}{4\pi^2\epsilon_0 m_\mathrm{X}^{2} c^3 r_\mathrm{C}^{2} E}~.
\end{equation}

\noindent
Here, $A_f$ is a charge-dependent amplification factor, $\epsilon_0$ is the vacuum permittivity, and $c$ is the speed of light in vacuum~\cite{fu1997quantum}.
For the colored CSL, Eq.~\ref{eq:radiation_rate_nmp} would be modified with a frequency-dependent non-trivial colored noise spectrum~\cite{carlesso2018colored}, which is beyond the scope of this letter.
The X-ray emission rate was searched for using Ge atoms in previous literature, first considering emissions from four outermost electrons~\cite{fu1997quantum} and later with the assumption that 30 quasi-free electrons would contribute to the radiation~\cite{piscicchia2017igex}. 
The amplification factor, which is $q^2$ for a single particle with charge $q$, is considered as $A_f = N_\mathrm{Ge}N_e e^2$ for a unit detector mass.
Here, $N_\textrm{Ge}$ is the number of germanium atoms per unit mass, $N_e=30$ is the number of quasi-free electrons in Ge, and $e$ is the elementary charge. 

In the m-p CSL, the X-ray emission rate is independent of $m_\mathrm{X}^2$~\cite{fu1997quantum}. 
In this case, Eq.~\ref{eq:radiation_rate_nmp} should be multiplied by a factor $(m_\mathrm{X}/m_0)^2$ where the reference mass $m_0$ is set as equal to the nucleon mass:

\begin{equation}\label{eq:radiation_rate_mp}
    \frac{d\Gamma(E)}{dE} = A_f \times \frac{\hbar \lambda}{4\pi^2\epsilon_0 m_0^{2} c^3 r_\mathrm{C}^{2} E}~.
\end{equation}

\noindent
The same amplification factor  $A_f = N_\mathrm{Ge}N_e e^2$ can be used considering emissions from quasi-free electrons~\cite{piscicchia2017igex}.

More stringent limits on $\lambda$ for the m-p CSL may be derived by considering the coherent X-ray emission from nuclei~\cite{donadi2021novel}. 
Here, the wavelength of the emitted photon $\lambda_k$ is smaller than the distance between electrons and the nucleus~($0.1$~nm) but much larger than the typical size of the nucleus~(10$^{-5}$~nm).
This corresponds to the energy range of $10-10^{5}$~keV~\cite{donadi2021novel}. 
In this case, the nucleus can be viewed as a single charged particle with $q=Ze$, where $Z=32$ is the atomic number of germanium.
In the m-p CSL where the contribution from nucleus is not suppressed by its large mass, the amplification factor becomes $A_f = N_\mathrm{Ge}\times(Z^2+30)\times e^2$~\cite{donadi2021novel}.
The $Z^2$ term is due to the coherent emission from nuclei, while 30 is from the quasi-free electrons.
In this letter, we consider both quasi-free electron-only assumption and the coherent nuclear emission model separately.  

The \MJ{} \DEM{}, described in detail in Ref.~\cite{abgrall2014majorana}, was designed to search for the neutrinoless double beta decay~($0\nu\beta\beta$) of $^{76}$Ge using arrays of \textit{p}-type point contact    high-purity germanium~(HPGe) detectors~\cite{abgrall2014majorana}. 
It consisted of two modules of HPGe detectors with a total mass of 44.1~kg, of which 29.7~kg were enriched to 88\% $^{76}$Ge, that operated from 2015 to 2021. 
With its ultralow background and unprecedented energy resolution~($\sigma=0.13$~keV at 10.37~keV) among experiments at a similar mass scale, the \MJ{} low energy program has demonstrated its potential to search for anomalous X-ray signatures from beyond Standard Model~(BSM) physics~\cite{peaksearch2017}.
Since the last release~\cite{peaksearch2017}, the collaboration's low energy program has improved its analysis tools, in both the background reduction and the efficiency determination~\cite{mjd_lat2022}. 
The achieved background level is 0.01~counts/(keV~kg~d) at 20~keV, and the efficiency is ($92.4\pm1.5)$\% at 20~keV. 
The exposure used for this analysis was collected between May~2016 and Nov.~2019, and reached 37.5~kg-y enriched exposure~\cite{mjd_lat2022}.

The theoretical lower bounds on $\lambda$ arise from the postulate that the CSL should resolve the measurement problem~\cite{Toro__2018_calculation}.
Imposing the condition that the collapse should be observable to the human eye at macroscopic scale, Ghirardi, Rimini and Weber suggested typical values of $\lambda = 10^{-16}$/s at a conventional value of $r_\mathrm{C} = 10^{-7}$~m~\cite{ghirardi1986,ghirardi1987,ghirardi1990old}. 
Bassi \textit{et al.} proposed $\lambda = 10^{-10\pm2}$/s~\cite{Bassi_2010_constraint} with the same postulate.
Adler suggested $\lambda = 10^{-8\pm2}$/s at $r_\mathrm{C}=10^{-7}$~m and $\lambda = 10^{-6\pm2}$/s at $r_\mathrm{C}=10^{-6}$~m from observations of the wavefunction collapse at mesoscopic scales~\cite{Adler_2007_bound}.
Variations of the theory permit a wide range of values and placing limits on the ($\lambda,r_\mathrm{C}$) parameter space test elements of the theory. 

We fit the model spectrum to the data using the unbinned extended likelihood method. 
In general, the efficiency-corrected spectral model $T(E)$ used to fit the data can be written as

\begin{equation}\label{eq:spectrum_model}
    T(E) = \epsilon(E) \times \Big[sW(E) + \sum_i b_{i}B_i(E) \Big]~,
\end{equation}

\noindent
where $\epsilon(E)$~is the detection efficiency for a single-sited X-ray event with energy~$E$, $s$~is the number of CSL-induced events, $W(E)$~is the normalized, $1/E$~dependent CSL PDF, $b_i$~is the number of events induced by the background~$i$, and $B_i$~is the normalized PDF of the background~$i$. 
The efficiency and the energy resolution were obtained from dedicated studies described in Ref.~\cite{mjd_lat2022}. 
All fit parameters were left free, except the positions and the widths of the known background peaks. 
The positions of the peaks were fixed at known values, while Gaussian constraints were imposed for the widths.
The energy resolutions at the peak energies were used as the mean of the constraints, with 30\% uncertainties. 
Since the CSL radiation signature extends across a broad fit range, the constraints on the peak widths have minimal impact on the sensitivity of this study.
No other constraints were imposed on the fit parameters.

The \MJ{} collaboration has studied the 5--100~keV region extensively, releasing multiple physics results~\cite{peaksearch2017,Wiseman_2020} over time. 
In this region, the background model consists of the tritium beta decay spectrum, a flat Compton continuum, the 46.5~keV $^{210}$Pb spectral line and an additional continuum associated with it~(hereafter mass-210 continuum), and known X-ray peaks~\cite{mjd_lat2022}. 
Contributions from other $\beta$-decaying isotopes such as $^{60}$Co and $^{63}$Ni are negligible compared to the Compton contribution with their considerably high endpoint energy~\cite{cdmslite2019cosmogenic}.

We set the fit range to be $19-100$~keV.
The upper bound of the fit range at 100~keV is chosen to fully utilize the entire \MJ{} low energy data to minimize the uncertainty of the flat spectra. 
As the X-ray signature expected from the CSL model is $1/E$-dependent, constraining the flat spectra enhances the sensitivity of this analysis. 
Above 100~keV, multi-sited events starts to be significant and the low energy cut efficiencies becomes unreliable. 
The lower bound of the fit range is set at 19~keV, to fully avoid the contribution from the tritium spectrum and minimize the contribution from the mass-210 continuum which is yet to be explored~\cite{mjd_lat2022}. 
As the mass-210 continuum has a rising tail at lower energies, we set the most conservative upper limit on the $1/E$ CSL radiation signature by limiting the fit range above 19~keV and assuming the mass-210 spectra to be flat.
The 30~outermost electrons may be treated as quasi-free, as the binding energy of the 2s orbital electrons in Ge~(1.414~keV) is more than one order of magnitude lower than 19~keV~\cite{piscicchia2017igex}.
The conditions for the coherent emission from nuclei discussed above is also fulfilled in this energy range.

The highest-energy non-negligible X-ray peak in the \MJ{} low energy spectrum is the 10.4~keV emission line from $^{68}$Ge~\cite{peaksearch2017}. 
No X-ray peak is observable within the fit range of $19-100$~keV~\cite{peaksearch2017,Wiseman_2020,avignone1992cosmogenic,genius2002cosmogenic,barabanov2006cosmogenic,mei2009cosmogenic,cebrian2010cosmogenic,zhang2016cosmogenic,edelweiss2017cosmogenic,wei2017cosmogenic,cdex2019cosmogenic,cdmslite2019cosmogenic}.
Eq.~\ref{eq:spectrum_model} can be written as

\begin{equation} \label{eq:spectrum_model2}
    T(E) = \epsilon(E) \times \Big[N_w\frac{s}{E} + N_gB_g(E) + N_fB_f + N_m\Theta(E-46.5) \Big]~, \\
\end{equation}

\noindent 
$w, g, f$ and $m$ represent the CSL signature, the $^{210}$Pb $\gamma$ Gaussian spectral line at 46.5~keV, the flat continuum, and the mass-210 continuum respectively. 
$B_f$ is a uniform distribution which extends throughout the entire fit range, while $\Theta(E-46.5)$ is the Heaviside step function with a cutoff at 46.5~keV.
$N_w=1/$(log(100)-log(19)), $N_f=(1/(100-19))$~keV$^{-1}$ and  $N_m=(1/(46.5-19))$~keV$^{-1}$ are normalization constants of corresponding PDFs.
The rate of wavefunction collapse per exposure, $R_0$, is given by

\begin{equation}
    R_0 = s\times\frac{N_w}{37.5}~/\textrm{(kg-y)}~.
\end{equation}

\begin{figure}[t]
    \centering
    \includegraphics[width=\columnwidth]{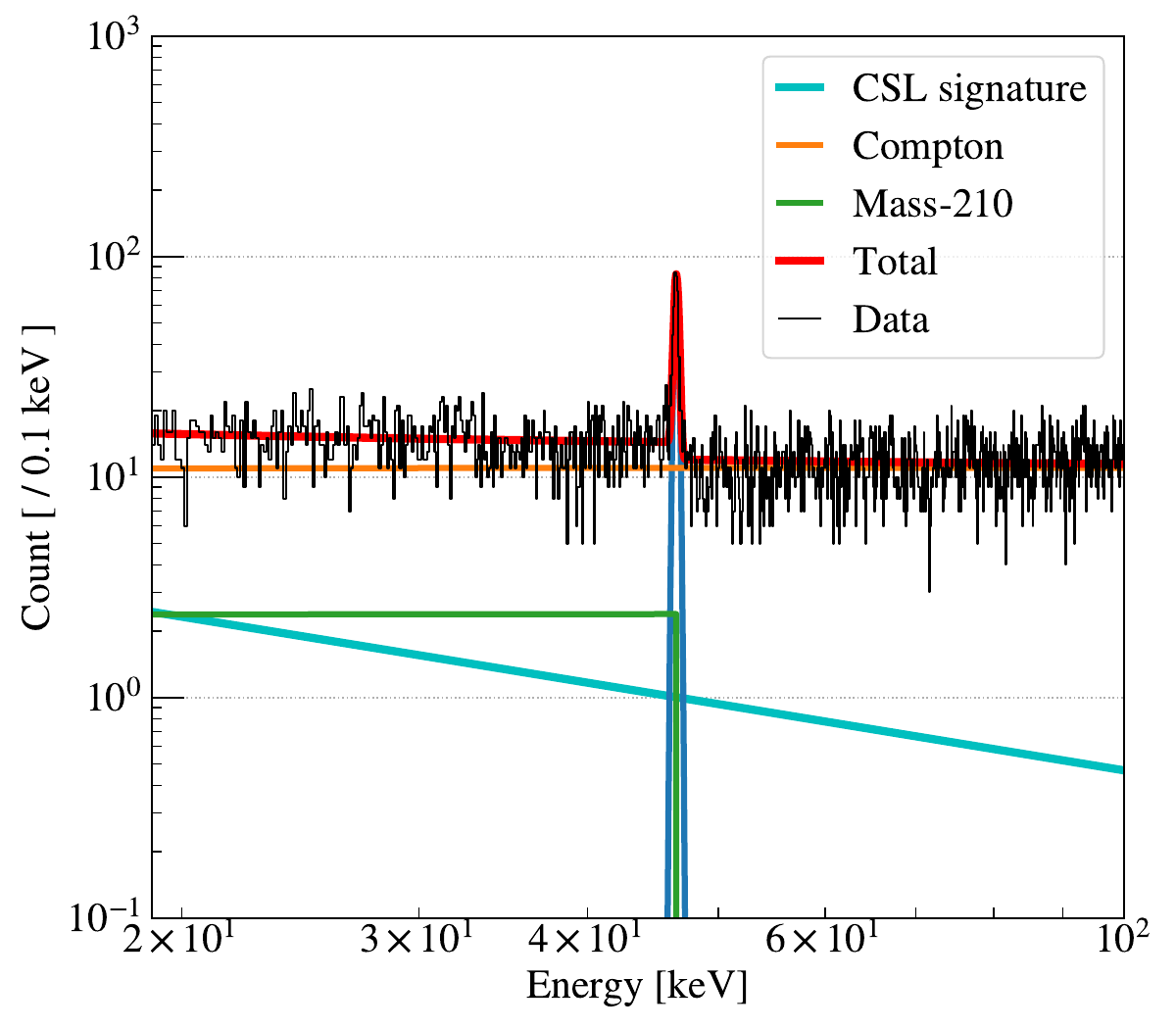}
    \caption{The 95\% CL spectral fit for the CSL radiation signature from wavefunction collapse. The 95\% CL upper limit value for $R_0$ is $0.0368$~/(kg-d).}
    \label{fig:bestfit}
\end{figure}

The best-fit value for the CSL signature with $1\sigma$ uncertainty is $s=(6.259\pm398.8)$ counts in the region from 19--100~keV region over a 37.5~kg-y exposure, which corresponds to $R_0 = (0.10\pm6.39)$ /(kg-y), or $(0.28\pm1.75)\times10^{-3}$ /(kg-d). 
We set an upper limit on the CSL signature by using the CLs method~\cite{Read_2002_CLs} with the $p$-value $p_\textrm{cls} = p_0/p_1$. 
The 95\% CL upper limit on $R_0$ is $0.0368$/(kg-d). 
Fig.~\ref{fig:bestfit} illustrates the 95\% CL upper limit.

The systematic uncertainties on this limit arise from energy determination and detection efficiency.
The energy parameter used for this analysis is determined by weekly calibration and was thoroughly studied in Ref.~\cite{ecal2022} and cross-checked specifically for the low energy analysis using the known X-ray peaks in Ref.~\cite{mjd_lat2022}.
The reported offset in the energy calibration is within 1\%. 
By adding small offsets within the bounds determined by this study of up to $\pm0.1$~keV to the energy of the data (and the efficiency curve, also from the data) we can estimate the impact of the uncertainty of the energy offset. 
The study found the maximum variation in the upper limit to be 0.3\%.
This analysis uses the final efficiency obtained from a dedicated study, which includes an uncertainty band around the nominal curve~\cite{mjd_lat2022}. 
If the lower bound efficiency is used instead of the nominal efficiency curve, the 95\% CL limit is weakened by 3.0\%. 
On the other hand, if the upper bound efficiency is considered, the limit improves by 3.1\%. 
Hence, we assign $\pm3.1$\%  systematic uncertainty on the final limit from the efficiency calculation.
The systematic uncertainties from the energy determination~(0.3\%) and the efficiency calculation~(3.1\%) are added in quadrature to get a 3.1\% total systematic uncertainty. 
For the nominal value of  $R_0 < 0.0368$~/(kg-d), the 3.1\% systematic uncertainty corresponds to $\pm 0.0011$~/(kg-d). 
Hence, the final 95\% CL limit for the \MJ{} CSL X-ray radiation signature $R_0 < (0.0368\pm 0.0011)$~/(kg-d).

\begin{figure}[t]
 \centering
 \includegraphics[width=\columnwidth]{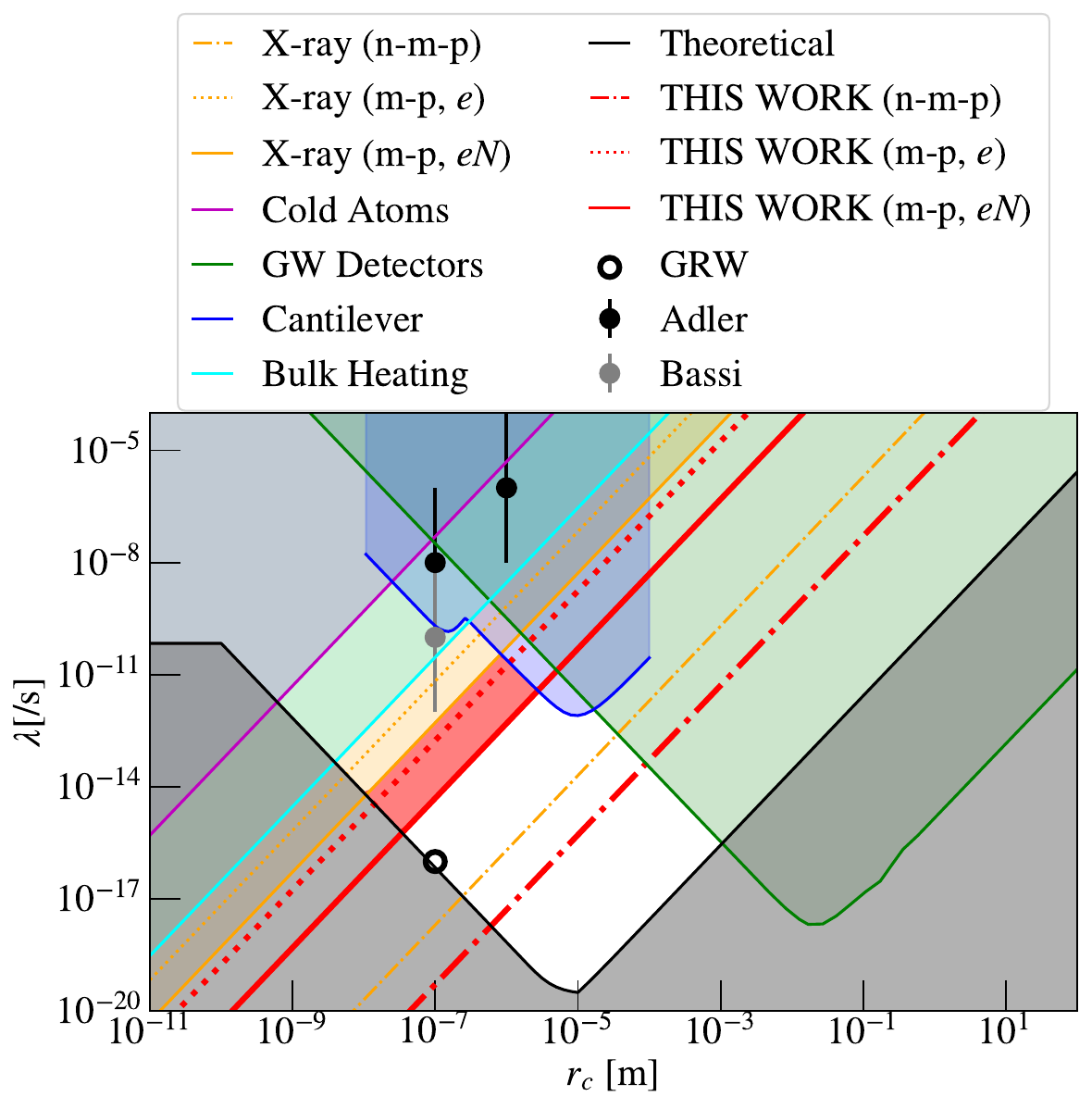}
 \caption{Upper limits on the wavefunction collapse parameters compared to existing upper limits from other experiments. Other limits are from cold atom experiments~(magenta)~\cite{BILARDELLO2016764,bilardello2017ultracold}, interpretation of the LIGO and LISA pathfinder gravity wave experiments~(green)~\cite{carlesso2016gravitational}, cantilever measurements~\cite{vinante2016nanocantilever,vinante2017improved,vinante2020layered} where an anomalous result has been reported~(blue)~\cite{vinante2017improved}, and CSL heating effect interpretation of heat leaks in low temperature experiments~(cyan)~\cite{adler2018bulk,bahrami2018bulk}. 
 The limits found from previous X-ray studies~\cite{piscicchia2017igex,donadi2021novel} are shown in orange lines.
 There is also a theoretical lower limit obtained by imposing that a graphene disk of radius 10~$\mu$m~(minimum resolution of human eye) is localized in less than 1~ms~(minimum time resolution of human eye)~(black)~\cite{Toro__2018_calculation}.
 Some proposed lower bounds by Adler~\cite{Adler_2007_bound}, Bassi~\cite{Bassi_2010_constraint}, and GRW~\cite{ghirardi1986} are also shown as black vertical lines, a gray vertical line, and a black hollow circle, respectively. }
 \label{fig:LimitSummary}
\end{figure}

Finally, to check that our choice of energy range does not bias our result, we varied the fit range between 18.7~keV and 20.0~keV at the lower bound and 95~keV and 100~keV at the upper bound.
The variation of the lower bound resulted in a maximum 5.4\% change on upper limit on $R_0$, which is consistent with the 3.1\% systematic uncertainty.
Varying the upper bound to lower energy reduced the achieved upper limit by up to 17\%, due to increased uncertainty in the flat background. 

We present this result in three different interpretations: 1) n-m-p limit where the contribution from nuclei is negligible, 2) m-p considering only the radiation from quasi-free electrons, and 3) m-p considering coherent emission from nuclei.
For the n-m-p CSL, Eq.~\ref{eq:radiation_rate_nmp} yields the limit in the $(\lambda-r_\mathrm{C})$ parameter space to be a band of $\lambda/r_\mathrm{C}^2 = (5.15\pm0.16)\times10^{-6}$ s$^{-1}$m$^{-2}$, which is a factor of 39 improvement over the previous limit~\cite{piscicchia2017igex}.
The result is presented in Fig~\ref{fig:LimitSummary} as a red dash-dotted line, with limits from previous experiments searching for the CSL signature for comparison. 
The bands from the systematic uncertainty are hidden within the line widths and not visible in this log-log plot.
We set the most stringent limit below $r_\mathrm{C}<10^{-4}$~m.

For the m-p CSL, Eq.~\ref{eq:radiation_rate_mp} yields

\begin{equation}
    \frac{\lambda}{r_\mathrm{C}^2} < (4.78\pm0.15)\times10^{18}\times \frac{m_0^2}{A_f}~[\mathrm{s}^{-1}\mathrm{m}^{-2}]~.
\end{equation}

\noindent
When only the 30 quasi-free electrons are considered to emit the X-ray radiation, the 95\% CL upper limit is a band of $\lambda/r_\mathrm{C}^2 = (17.4\pm0.5)$ s$^{-1}$m$^{-2}$.
This is also a factor of 39 improvement over the previous best limit~\cite{piscicchia2017igex}.
Our result is the first to fully exclude the theoretical value suggested by Bassi \textit{et al.} with this assumption.
If the coherent emission from nuclei is considered, the 95\% CL limit for the m-p CSL is a band of $\lambda/r_\mathrm{C}^2 = (4.94\pm0.15)\times10^{-1}$ s$^{-1}$m$^{-2}$ in the $(\lambda-r_\mathrm{C})$ parameter space.
This is a factor of 105 improvement over the previous best limit~\cite{donadi2021novel} with the same assumption, and we set the most stringent limits in the region $r_\mathrm{C}<10^{-6}$~m.
The two limits are shown in Fig.~\ref{fig:LimitSummary} as a red dotted line and a red solid line, respectively.

Using the same result, we present a similar limit for another version of the collapse model, the Di\`{o}si-Penrose~(DP) model~\cite{diosi1989models,penrose1996gravity} which has been extensively studied in recent literature~\cite{donadi2021underground}. 
Similarly to the CSL, the DP model predicts a radiation signature with an energy dependence of $1/E$. 
Hence the same fit can place a limit on the DP model. 

The rate of the DP radiation signature is given in Ref.~\cite{donadi2021underground} as 

\begin{equation}\label{eq:radiation_rate_DP}
    \frac{d\Gamma(E)}{dE} = \frac{2}{3}\frac{e^2 G Z^2 N_\mathrm{Ge} M}{\pi^{3/2} \epsilon_0 c^3} \frac{1}{R_\textrm{DP}^3}\frac{1}{E}~,
\end{equation}

\noindent
where the physical constants $G$ is the gravitational constant, $Z$ is the atomic number of germanium, $M$ is the mass of the experiment, and $R_\textrm{DP}$ is the characteristic cut-off length of the DP model~\cite{donadi2021underground}. 

With the upper limit of $R_0 = (0.0368\pm 0.0011)$~/(kg-d), we get a 95\% CL lower bound on the cut-off length of  $R_\mathrm{DP} > (2.54\pm0.03)\times 10^{-10}$~m.
This is more stringent than the previous best limit of $5.4\times 10^{-11}$~m~\cite{donadi2021underground} and more than an order of magnitude higher than  $R_\mathrm{DP} > 5.0\times 10^{-12}$~m predicted by Penrose for germanium crystals~\cite{donadi2021underground}. 

We analyzed the \MJ{} low energy data in the 19-100~keV region to search for the X-ray radiation signature from the CSL wavefunction collapse model. 
We investigated the non-mass-proportional and the mass-proportional versions of the CSL with two different assumptions, first by considering only the emissions from the quasi-free electrons and then considering the coherent emission from nuclei. 
Our limits on the white CSL are orders of magnitude lower than the former leading X-ray limits, and are the most stringent in the field for a range of $r_\mathrm{C}$ values $<10^{-6}$~m depending on models.
If only the quasi-free electrons are considered to emit the X-ray radiation, our result is the first to exclude the theoretical value suggested by Bassi \textit{et al.} for the m-p CSL. 
Extensions of the CSL models such as the colored noise models are required to evade such stringent constraints. 
In these terms, the \MJ{} result can be interpreted as a test of the white noise model itself.
Our result experimentally motivates the pursuit of the colored CSL with non-trivial frequency spectrum. 
We also improved the lower limit on the Di\`{o}si-Penrose collapse model by almost an order of magnitude, adding significance to the exclusion of the parameter-free version of the model.

This material is based upon work supported by the U.S.~Department of Energy, Office of Science, Office of Nuclear Physics under contract / award numbers DE-AC02-05CH11231, DE-AC05-00OR22725, DE-AC05-76RL0130, DE-FG02-97ER41020, DE-FG02-97ER41033, DE-FG02-97ER41041, DE-SC0012612, DE-SC0014445, DE-SC0018060, and LANLEM77/LANLEM78. We acknowledge support from the Particle Astrophysics Program and Nuclear Physics Program of the National Science Foundation through grant numbers MRI-0923142, PHY-1003399, PHY-1102292, PHY-1206314, PHY-1614611, PHY-1812409, PHY-1812356, and PHY-2111140. We gratefully acknowledge the support of the Laboratory Directed Research \& Development (LDRD) program at Lawrence Berkeley National Laboratory for this work. We gratefully acknowledge the support of the U.S.~Department of Energy through the Los Alamos National Laboratory LDRD Program and through the Pacific Northwest National Laboratory LDRD Program for this work.  We gratefully acknowledge the support of the South Dakota Board of Regents Competitive Research Grant. 
We acknowledge the support of the Natural Sciences and Engineering Research Council of Canada, funding reference number SAPIN-2017-00023, and from the Canada Foundation for Innovation John R.~Evans Leaders Fund.  This research used resources provided by the Oak Ridge Leadership Computing Facility at Oak Ridge National Laboratory and by the National Energy Research Scientific Computing Center, a U.S.~Department of Energy Office of Science User Facility located at Lawrence Berkeley National Laboratory. We thank our hosts and colleagues at the Sanford Underground Research Facility for their support.

\bibliography{main}

\end{document}